\documentclass[preprints,article,accept,moreauthors,pdftex]{Definitions/mdpi} 

\firstpage{1} 
\pubvolume{1}
\issuenum{1}
\articlenumber{0}
\pubyear{2021}
\copyrightyear{2020}
\datereceived{} 
\dateaccepted{} 
\datepublished{} 
\hreflink{https://doi.org/} 

\usepackage{bm}
\usepackage{braket}

\Title{Effect of interlayer coupling and symmetry on high-order harmonic generation from monolayer and bilayer hexagonal boron nitride}

\TitleCitation{Title}

\Author{Dasol Kim $^{1, 2, \dagger}$, Yeon Lee $^{1, 2, \dagger}$, Alexis Chac\'{o}n $^{1, 2}$ and Dong Eon Kim $^{1, 2}$* }

\AuthorNames{Dasol Kim, Yeon Lee, Dong Eon Kim and Alexis Chac\'{o}n}

\AuthorCitation{Dasol, K.; Yeon, L.; Dong Eon, K.; Alexis, C.}

\address{%
$^{1}$ \quad Department of Physics and Center for Attosecond Science and Technology, POSTECH, 7 Pohang 37673, South Korea; dsestil@gmail.com (D. Kim), dldusigo@gmail.com (Y. Lee), achacon@postech.ac.kr (A. Chac\'on)\\
$^{2}$ \quad Max Planck POSTECH/KOREA Research Initiative, Pohang 37673, South Korea; 
}

\corres{Correspondence: kimd@postech.ac.kr (D. E. Kim)}

\firstnote{These authors contributed equally to this work.} 

\abstract{
High-order harmonic generation (HHG) is a fundamental process which can be simplified as the production of high energetic photons from a material subjected to a strong driving laser field.
This highly nonlinear optical process contains rich information concerning the electron structure and dynamics of matter, for instance, gases, solids and liquids. 
Moreover, the HHG from solids has recently attracted the attention of both attosecond science and condensed matter physicist, since the HHG spectra can carry information of electron-hole dynamics in bands and inter- and intra-band current dynamics.
In this paper, we study the effect of interlayer coupling and symmetry in two-dimensional (2D) material by analyzing high-order harmonic generation from monolayer and two differently stacked bilayer hexagonal boron nitrides (hBNs). 
These simulations reveal that high-order harmonic emission patterns strongly depend on crystal inversion symmetry (IS), rotation symmetry and interlayer coupling.}


\begin{document}

\section{Introduction}\label{sec:intro}

Thanks to the significant advances in laser technology within the visible, infrared and terahertz regions, there has been a remarkable interest in understanding the nonlinear process in materials under the influence of a strong laser field. This is important not only for strong-field physics in condensed matter~\cite{vampa2015a, you2017, tancogne-dejean2017, luu2018, kwon2016, langer2018} but also for new technologies such as ultrafast optoelectronic and photonic applications. 

One of the most representative nonlinear processes in solid-state subjected to a strong laser field is high-order harmonic generation (HHG). Initially, HHG has been extensively studied and used in gases and molecules for table-top XUV sources as well as a tool to explore ultrafast electron dynamics~\cite{krausz2009}. Recently, research on HHG has been extended to solid state physics into the materials such as ZnO, MgO, GaSe, graphene, GaP meta surface, transition metal dichalcogenides (TMDCs), Cd$_3$As$_2$, hexagonal boron nitride (hBN) and topological materials~\cite{shambhu2019, shambhu2011, you2017, you2017b, Liu2017, Shcherbakov2021, yoshikawa2017, kovalev2020, tancogne-dejean2018, yue2020_2, baykusheva2021}. The HHG in solid-state materials may provide more compact and efficient XUV sources as well as the information on electron structures and ultrafast electron-hole dynamics~\cite{vampa2015b, itatani2004}.

In recent years, two-dimensional (2D) materials have been studied for potential applications to next-generation nano-photonics and nano-electronics devices~\cite{tancogne-dejean2017a, langer2017, hammond2017, sivis2017, garg2018}. 2D materials are unique in terms of the potential to combine various atomically thin layers to control material properties or even create new materials such as so-called van der Waals heterostructure~\cite{geim2013}. Since van der Waals interaction between layers is rather weak, 2D materials with different compositions can be readily assembled, providing additional knobs to manipulate material properties, such as conductivity or semi-metal characteristics~\cite{liao2018, kang2020, reddy2020}. 

Nowadays the stacking of atomically thin layers with a high degree of control has become a reality~\cite{novoselov2016, yankowitz2018, tsen2016, dean2010}. Layer spacing can even be controlled by pressure~\cite{chen2020} or doping~\cite{ding2021}. It is therefore timely and demanding to understand how stacking scheme and interlayer coupling affect the optical properties of HHG in the context of strong-field physics. 
The identification of clear fingerprints related to a particular stacking or layer coupling could lead to HHG spectroscopy for monolayer, bilayer or few-layer systems. 

Because of the high damage threshold, hBN is a promising candidate for light emission in the far UV region via HHG~\cite{bourrellier2014, bourrellier2016}. Theoretical investigations on the various characteristics of HHG from hBN, such as rotation anisotropy, time-frequency behavior and wavelength scaling of harmonic yields, have been pursued recently~\cite{tancogne-dejean2018, yu2018, kong2021}. Improvements in growth technology have enabled hBN to be stacked in various ways~\cite{kim2013, khan2017, shi2020}. Pioneering research has indicated that among five high symmetry stacking orders, the AB and AA$'$ stacking orders are the most stable configurations in either bilayer systems or bulk~\cite{ribeiro2011, constantinescu2013, gilbert2019}.

In the present paper, using well-established semiconductor Bloch equations (SBEs) with a tight-binding approach to model the effect of interlayer coupling on the HHG process, we explore the HHG features of the monolayer and bilayer hBNs driven by linearly, circularly or elliptically polarized laser fields. 

This paper is organized as follows: the methodology is discussed in Section~\ref{sec:methods}. In Section~\ref{sec:results}, we present the calculated HHG spectra in the monolayer, AB and AA$'$ bilayers by varying laser polarization angle, ellipticity of driving laser field and interlayer coupling. The HHG spectra are analyzed in terms of symmetry and interlayer coupling. Finally, we summarize our findings and analysis in Section~\ref{sec:conclustion}. 


\section{Methods}\label{sec:methods}

\end{paracol}
\begin{figure}[H]
    \widefigure
    \centering
    \includegraphics{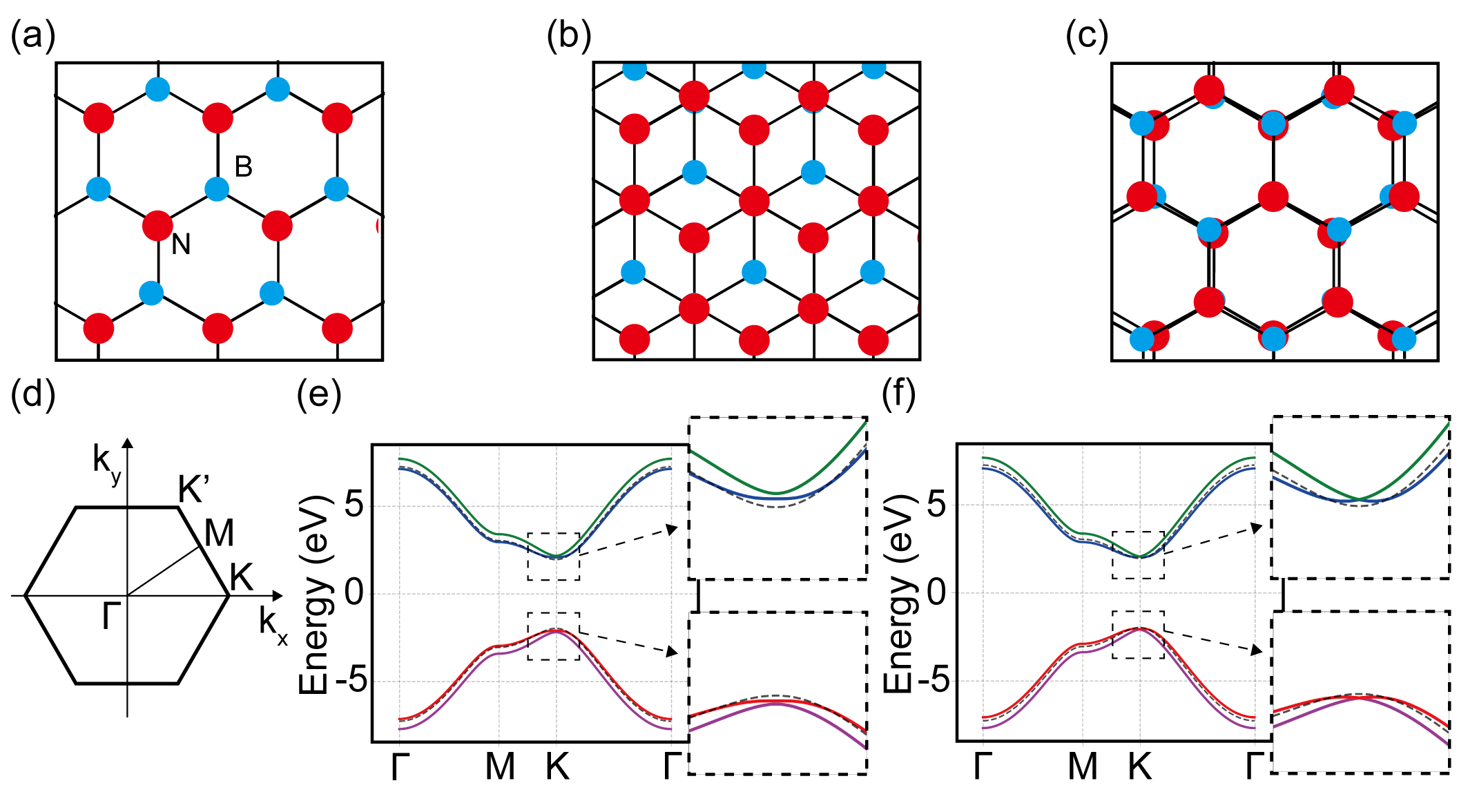}
    \caption{Top view of the (a) monolayer, (b) AB bilayer and (c) AA$'$ bilayer hexagonal boron nitrides (hBNs). B atoms are indicated in blue and N atoms are depicted in red. (d) Schematic Brillouin zone of hBN. The x-axis is along the $\Gamma-K$ direction, while the y-axis is along the $\Gamma-M$ direction. (e) The energy band dispersion of the monolayer and AB bilayer hBN. The bands from the monolayer are displayed with black dashed lines and those of the bilayer are represented by colored solid lines (CB2-green; CB1-blue, VB2-red, VB1-magenta). (f) The energy dispersion is displayed in the same way in (e) for the monolayer and AA$'$ bilayer hBN. Band crossing is observed at $K$-point in valence and conduction bands of the AA$'$ bilayer.}
    \label{fig:material_structure}
\end{figure}
\begin{paracol}{2}
\switchcolumn

Figure~\ref{fig:material_structure}(a) shows a top view of the monolayer hBN, which has unit cell parameter a=b=2.5 {\AA}~\cite{han2008}, in real space. The Monolayer hBN belongs to the $D_{3h}$ point group. It has both 3-fold symmetry and broken inversion symmetry (IS). Figures~\ref{fig:material_structure}(b-c) are the real space representation of the AB and AA$'$ stackings, respectively. Both stacking types share the same interlayer distance, 3.3 {\AA}~\cite{han2008}. The corresponding Brillouin zone (BZ) is visualized in Fig.~\ref{fig:material_structure}(d). We choose the $\Gamma-K$ direction as x direction and the $\Gamma-M$ direction as y direction.
Figures~\ref{fig:material_structure}(e-f) are the energy dispersion relation of the AB and AA$'$ stacking in comparison with the monolayer case. We choose the same stacking notation (AB or AA$'$) as done in the work by Ribeiro and Peres~\cite{ribeiro2011}. These two different stacking schemes are chosen to emphasize the symmetry change between the monolayer and bilayers – the symmetry of the AB bilayer is almost the same as the symmetry of the monolayer (belonging to point group $C_{3v}$), while the AA$'$ bilayer exhibits a huge difference (point group $D_{3d}$). In detail, the monolayer has x-y plane as a mirror plane, but the AB bilayer does not have such a mirror plane, i.e. $H(z) \neq H(-z)$. This is the only difference between the monolayer and AB bilayer and it is related to symmetry in z-direction. In this paper, we study 2D material, hence $\bm{k}_{z}$ is not a good quantum number. Moreover, we propagate a laser into the material in the surface normal direction, and do not consider applying an out-of-plane electric field. Hence, we can assume that the monolayer and AB bilayer have the same symmetry configuration in this paper. 

Unlike the other two cases, the AA$'$ bilayer recovers IS, and this makes a large difference between the AA$'$ bilayer and the other two cases. It is also worth noting that the AA$'$ bilayer has degeneracy at $K$ point while the AB bilayer has band splitting as shown in the inset of Figure~\ref{fig:material_structure}(e,f).

\subsection{Hamiltonian of the hBN}~\label{sec:Hamiltonian}

We use a simple Hamiltonian that considers only one $p_{z}$ orbital (or hybridized orbital) from B atom and N atom. Only considering nearest neighbor (NN) atoms, a monolayer Hamiltonian is written as 

\begin{align}
    H_{0} = 
    \begin{bmatrix}
    M_{0} & \gamma  \\
    \gamma^{*} & -M_{0}
    \end{bmatrix}.
    \label{eq:Hamiltonian_monolayer}
\end{align}

\noindent On-site energy difference is written as $M_{0} = (\varepsilon_{B} - \varepsilon_{N})/2$, where $\varepsilon_{B}$ is on-site energy in the B atom and $\varepsilon_{N}$ is on-site energy in the N atom. The interaction between B and N atoms is

\begin{align}
    \gamma=t_{1}\sum_{i=1}^{3} e^{i\bm{k}\cdot\bm{a}_{i}},
    \label{eq:off_diagonal_mono_Hamiltonian}
\end{align}

\noindent where $t_{1}$ is the NN hopping parameter, and $\bm{a}_{i}$ is the NN vector from B atom to N atom~\cite{chacon2020}. The first row or column is corresponding to $p_{z}$ orbital at B atom ($p_{z}^{B}$) and the second refers to $p_{z}$ orbital at N atom ($p_{z}^{N}$). A bilayer Hamiltonian can be constructed by `stacking' the monolayer Hamiltonian, $H_0$. Before stacking, it is mandatory to carefully consider the lattice orientation. For example, in the AA$'$ bilayer in Figure~\ref{fig:material_structure}(c), the top layer is rotated 180$^{\circ}$ with respect to the bottom layer. This rotation makes $\bm{a}_{i} \xrightarrow{} -\bm{a}_{i}$, i.e. $H_{0} \xrightarrow{} H_{0}^{*}$ or $M_{0} \xrightarrow[]{} -M_{0}$ by exchanging B with N atom.

We can directly stack the monolayer Hamiltonian, $H_{0}$, to obtain the AB bilayer Hamiltonian because there is no rotation of the layers in AB stacking, as shown in Figure~\ref{fig:material_structure}(b).
Then, the AB bilayer Hamiltonian basis can be referred to $\{p_{z}^{B,(\mathrm{t})}, p_{z}^{N,(\mathrm{t})}, p_{z}^{B,(\mathrm{b})}, p_{z}^{N,(\mathrm{b})}\}$ where (t) is a superscript for the top layer and (b) refers to the bottom layer. For interlayer direction, we only include interlayer NN interaction between two vertically overlapped atoms - N atoms of the top layer and B atoms of the bottom layer (Figure~\ref{fig:material_structure}(b)). Then the Hamiltonian of the AB bilayer is 

\begin{align}
    H_{\mathrm{AB}} = 
    \begin{bmatrix}
    M_{0} & \gamma & 0 & 0 \\
    \gamma^{*} & -M_{0} & t_{11} & 0 \\
    0 & t_{11} & M_{0} & \gamma \\
    0 & 0 & \gamma^{*} & -M_{0}
    \end{bmatrix},
    \label{eq:Hamiltonian_AB}
\end{align}

\noindent where $t_{11}$ is the interlayer hopping parameter. It is worth noting that the interlayer NN vector has only z-component, hence there is no phase factor in interlayer terms. 

The same procedure can be applied to the AA$'$ bilayer. For the case of the AA$’$ bilayer, the top layer is inverted with respect to the bottom layer. \sloppy We choose $\{p_{z}^{B,(\mathrm{b})}, p_{z}^{N,(\mathrm{b})}, p_{z}^{N,(\mathrm{t})}, p_{z}^{B,(\mathrm{t})}\}$ as the set of atomic orbitals corresponding to the AA$'$ bilayer Hamiltonian basis. The interlayer NN interaction is in-between different atomic species that are directly stacked on top of each other (see Figure~\ref{fig:material_structure}(c)). The Hamiltonian of the AA$'$ bilayer is written as

\begin{align}
    H_{\mathrm{AA}'} = 
    \begin{bmatrix}
    M_{0} & \gamma & t_{11} & 0 \\
    \gamma^{*} & -M_{0} & 0 & t_{11} \\
    t_{11} & 0 & -M_{0} & \gamma \\
    0 & t_{11} & \gamma^{*} & M_{0}
    \end{bmatrix}.
    \label{eq:Hamiltonian_AA'}
\end{align}
 
The tight binding parameters from the study of Ribeiro and Peres~\cite{ribeiro2011} are given in Table~\ref{table:Hamiltonian_parameters}. The $M_0$ and $t_1$ values change slightly depending on the stacking type even if monolayer structure remains the same. This happens because the tight-binding Hamiltonian is constructed by projecting the result of DFT calculation. Some effects, such as interlayer next-nearest-neighbor (NNN) coupling or interlayer third-nearest-neighbor (NNNN) coupling, are neglected in the model of this paper. These neglected effects are projected into the monolayer Hamiltonian and interlayer NN hopping parameter. This indicates that the interlayer coupling partially contributes to the monolayer Hamiltonian.   

\begin{specialtable}[H]
\small
\caption{Tight-binding parameters for the monolayer, AB bilayer and AA$'$ bilayer hBN from~\cite{ribeiro2011}. Parameters are fitted to the valence band calculated by DFT. \label{table:Hamiltonian_parameters}}
\begin{center}
\begin{tabular}{cccc}
\toprule
\textbf{Stacking type}	& $M_{0}$ (eV)	& $t_{1}$ (eV) & $t_{11}$ (eV) \\
\midrule
monolayer& 1.96	&	2.33	&  -\\
AB		& 2.095	&	2.37	&   0.60\\
AA$'$   & 2.04  &   2.36    &   0.32\\
\bottomrule
\end{tabular}
\end{center}
\end{specialtable}

\subsection{Semiconductor Bloch equations in tight-binding basis} \label{sec:SBEs}

From the excellent work of Virk and Sipe~\cite{virk2007}, SBEs in length gauge (LG) are frequently used as a simulation framework to study the HHG process in solids. However, there is a problem related to the phase of Hamiltonian eigenstate~\cite{virk2007, yue2020}. An eigenstate of Hamiltonian has the freedom to choose phase, $\ket{\psi_{n,\bm{k}}} \rightarrow e^{i\phi}\ket{\psi_{n,\bm{k}}}$, without changing the property of the material. Even if this gauge transformation does not change the material characteristics, it affects the simulation result as a numerical error. Constructing a continuous, or good enough, wavefunction gauge to minimize numerical noise has been another big issue. To overcome this problem, we solve the SBEs in tight-binding basis~\cite{silva2019, kim2021} instead of eigenstate basis as in other papers~\cite{virk2007, golde2008, vampa2014, yue2020, chacon2020}. To distinguish the two different basis, we use superscript (W) to indicate the tight-binding basis gauge, and superscript (H) to indicate the eigenstate basis gauge as in~\cite{wang2006}.

Let us set localized Wannier functions $w_{n}(\bm{r}-\bm{R}) = \braket{\bm{r}|\bm{R}n}$, where $\bm{R}$ is Bravais lattice vector. We assume that Wannier functions are orthogonal, $\braket{\bm{R}_{m}|\bm{R}'_{n}} = \delta_{m,n}\delta_{\bm{R},\bm{R}'}$. Tight-binding basis is represented from the Wannier functions as Bloch-like basis

\begin{align}
    \ket{\psi^{(\mathrm{W})}(\bm{k})} = \frac{1}{\sqrt{N}}\sum_{\bm{R}}e^{i\bm{k}\cdot(\bm{R}+\bm{\Delta}_{n})}\ket{\bm{R}n},
    \label{eq:Wannier_basis}
\end{align}

\noindent where $N$ is the number of unit cells included as a normalization factor, and $\bm{\Delta}_{n}$ is the Wannier center ($\bm{\Delta}_{n}=\braket{\bm{0}n|\hat{\bm{r}}|\bm{0}n}$). For the model used in this paper, Wannier centers indicate the locations of the atoms in an unit cell, as they are localized around an atom. The components of the Hamiltonian matrix in tight-binding basis are

\begin{align}
    H_{mn}^{(\mathrm{W})}(\bm{k}) &= \braket{\psi^{(\mathrm{W})}_{m}(\bm{k})|\hat{H} | \psi^{(\mathrm{W})}_{n}(\bm{k}) } \nonumber \\
    &= \sum_{\bm{R}} e^{i\bm{k}\cdot(\bm{R} + \bm{\Delta}_{m}-\bm{\Delta}_{n})}\braket{\bm{0}_{m}|\hat{H}_{0} | \bm{R}_{n}}.
\label{eq:Hamiltonian_wannier}
\end{align}

\noindent By comparing Eq.~\ref{eq:off_diagonal_mono_Hamiltonian} and Eq.~\ref{eq:Hamiltonian_wannier}, it is clear that the Hamiltonians used in this paper are constructed from the basis defined in Eq.~\ref{eq:Wannier_basis}. The NN vector $\bm{a}_{i}$ in Eq.~\ref{eq:off_diagonal_mono_Hamiltonian} is the NN displacement between two atoms which can be written as $\bm{R}+\bm{\Delta}_{N}-\bm{\Delta}_{B}$. 

It is worth noting that our basis choice is a bit different from previous research ($\ket{\psi^{(\mathrm{W})}(\bm{k})} = \sum_{\bm{R}}e^{i\bm{k}\cdot\bm{R}}\ket{\bm{R}n}$)~\cite{wang2006, ribeiro2011, silva2019}. We choose Eq.~\ref{eq:Wannier_basis} as a basis to handle dipole transition matrix elements (DTMEs) in tight-binding basis. Well localized Wannier functions can be approximated as $\braket{\bm{R}m|\hat{\bm{r}}|\bm{0}n} = \bm{\Delta}_{n}\delta_{mn}$. With this so-called `diagonal' approximation and our basis choice in Eq.~\ref{eq:Wannier_basis}, we can eliminate DTMEs and Berry connection in tight-binding basis, $\braket{u^{(\mathrm{W})}_{m}(\bm{k})| \nabla_{\bm{k}} | u^{(\mathrm{W})}_{n}(\bm{k}) } = 0$, where $\ket{u^{(\mathrm{W})}_{m}(\bm{k})} = e^{-i\bm{k}\cdot\bm{r}}\ket{\psi^{(\mathrm{W})}_{m}(\bm{k})}$. This basis transformation only affects DTMEs or phase factor, $e^{i\bm{k}\cdot(\bm{R} + \bm{\Delta}_{m}-\bm{\Delta}_{n})}$ in Eq.~\ref{eq:Hamiltonian_wannier}, and do not change the terms like hopping parameters or on-site energies, $\braket{\bm{0}_{m}|\hat{H}_{0} | \bm{R}_{n}}$.

We calculate laser-matter interaction through the SBEs in LG with tight-binding basis~\cite{silva2019, kim2021},

\begin{align}
    i\frac{\partial}{\partial t}\rho^{(\mathrm{W})}(\bm{K}, t) = [H_{0}^{(\mathrm{W})}(\bm{K}+\bm{A}(t)), \rho^{(\mathrm{W})}(\bm{K}, t)], \label{eq:eom_density_matrix}
\end{align}

\noindent where $\bm{A}(t)$ is the vector potential of the laser, and $\rho^{(\mathrm{W})}$ is the density matrix in tight-binding basis whose components are $\rho^{(\mathrm{W})}_{mn} = \braket{\psi^{(\mathrm{W})}_{m}(\bm{k})|\hat{\rho} | \psi^{(\mathrm{W})}_{n}(\bm{k}) }$. $H_{0}^{(\mathrm{W})}$ is a tight-binding Hamiltonian as we discussed in Eq~\ref{eq:Hamiltonian_wannier}. As in previous papers~\cite{vampa2014, vampa2015, chacon2020}, we represent the dissipation term through a phenomenological constant, dephasing time $T_{2}$. Since we cannot apply the well-known decoherence term $\Gamma_{mn} = (1-\delta_{mn})/T_{2}$ directly to the
tight-binding basis, we handle the dephasing term in the eigenstate basis gauge. In the eigenstate basis, SBEs including the dissipation term are written as 

\begin{align}
    i\frac{\partial}{\partial t}\rho^{(\mathrm{H})}_{mn}(\bm{K}, t) &= [H_{0}^{(\mathrm{H})}(\bm{K}+\bm{A}(t)), \rho^{(\mathrm{H})}(\bm{K}, t)]_{mn} \nonumber \\
    &+ \bm{E}(t)\cdot[\bm{D}^{(\mathrm{H})}(\bm{K}+\bm{A}(t)),\rho^{(\mathrm{H})}(\bm{K},t)]_{mn} - i\frac{1-\delta_{mn}}{T_{2}}\rho^{(\mathrm{H})}_{mn}(\bm{K}, t).
    \label{eq:eom_density_matrix_Hamiltonian}
\end{align}

\noindent There are two additional terms in comparison with Eq.~\ref{eq:eom_density_matrix}. The first one is term which contains laser electric field $\bm{E}(t)$ and dipole matrix in eigenstate basis $\bm{D}^{(\mathrm{H})}$ whose diagonal components are the Berry connection and off-diagonal components are DTMEs. This term is omitted in Eq.~\ref{eq:eom_density_matrix}, since $\bm{D}^{(\mathrm{W})}=0$ in our approximation. The final term in Eq.~\ref{eq:off_diagonal_mono_Hamiltonian} is the dephasing term which comes from electron-electron, electron-phonon scattering~\cite{vampa2014} and propagation effects~\cite{yabana2012, floss2018}. Also, it is worth noting that $H_{0}^{(\mathrm{H})}$ is the Hamiltonian matrix in the eigenstate basis, hence the diagonal matrix of band energy, $(H_{0}^{(\mathrm{H})})_{mn} = \varepsilon_{m}\delta_{m,n}$. We handle the dephasing time in the eigenstate basis and calculate other terms in tight-binding basis. The gauge transformation is performed in every time step as in~\cite{silva2019,kim2021}.

The microscopic electric current is defined with a density matrix that reads:

\begin{align}
    \bm{j}(\bm{K}, t) &= -\mathrm{Tr} (\hat{v}\hat{\rho}) \nonumber \\
    &= -\sum_{m, n} \rho_{m,n}^{(\mathrm{W})}(\bm{K}, t)\bm{P}_{nm}^{(\mathrm{W})}(\bm{K}+\bm{A}(t)),
\end{align}

\noindent where $\bm{P}_{nm}^{(\mathrm{W})}(\bm{k}) = \braket{\psi_{n}^{(\mathrm{W})}|\partial_{\bm{k}}\hat{H}_{0}^{(\mathrm{W})}|\psi_{m}^{(\mathrm{W})}}$~\cite{aversa1995, kruchinin2013, yue2020, silva2019}. We integrate this over the first BZ, and calculate the macroscopic currents

\begin{align}
    \bm{J}(t) = \int_{\mathrm{BZ}}\frac{d\bm{k}}{(2\pi)^{2}} \bm{j}(\bm{K}, t).
\end{align}

\noindent Finally, the intensity of the HHG spectrum is calculated by Fourier transforming the current  $\bm{J}(t)$:

\begin{align}
    I(w) = \left|\mathcal{F}\left[\frac{d}{dt}\bm{J}(t)\right]\right|^{2}.
\end{align}

In our calculation, we use a laser with the vector potential

\begin{align}
    \bm{A}\left(t\right)= \frac{E_0}{w_0}f(t)\left(-\frac{1}{\sqrt{1+\epsilon^2}}\sin{\left(w_{0}(t-t_{0})-\phi_{0}\right)}\hat{\bm{e}_{x}}+\frac{\epsilon}{\sqrt{1+\epsilon^2}}\cos{\left(w_{0}(t-t_{0})-\phi_{0}\right)}\hat{\bm{e}_{y}}\right),
    \label{eq:vecotr_potential}
\end{align}

\noindent where $E_{0}$ is the electric field peak strength, $w_{0}$ is laser angular frequency, $\phi_{0}$ is the carrier-envelope phase (CEP), $\epsilon$ is the ellipticity of the laser, and $f(t)$ is the laser pulse envelope. We use Gaussian envelope $f(t) = \exp[-4\log{2}\frac{(t-t_{0})^{2}}{t_{\sigma}^{2}}]$ where $t_{0}$ is the center of the pulse and $t_{\sigma}$ is the full width at half maximum (FWHM) of the laser. The electric field is given by $\bm{E}(t) = -\partial_{t} \bm{A}(t)$. Eq.~\ref{eq:vecotr_potential} is constructed in a way that the electric field has its maximum when $t=t_{0}$ and $\phi_{0}=0$ at least in a slow varying envelope. We use laser parameter $E_{0}$ = 0.3 V/\AA, $w_{0}$ = 0.02398 a.u. (corresponding to wavelength $\lambda$ = 1.9 $\mu$m), and FWHM of 10 cycles with a Gaussian envelope.

\section{Results and Discussion}\label{sec:results}

\begin{figure}[H]
    \centering
    \includegraphics[width=14cm]{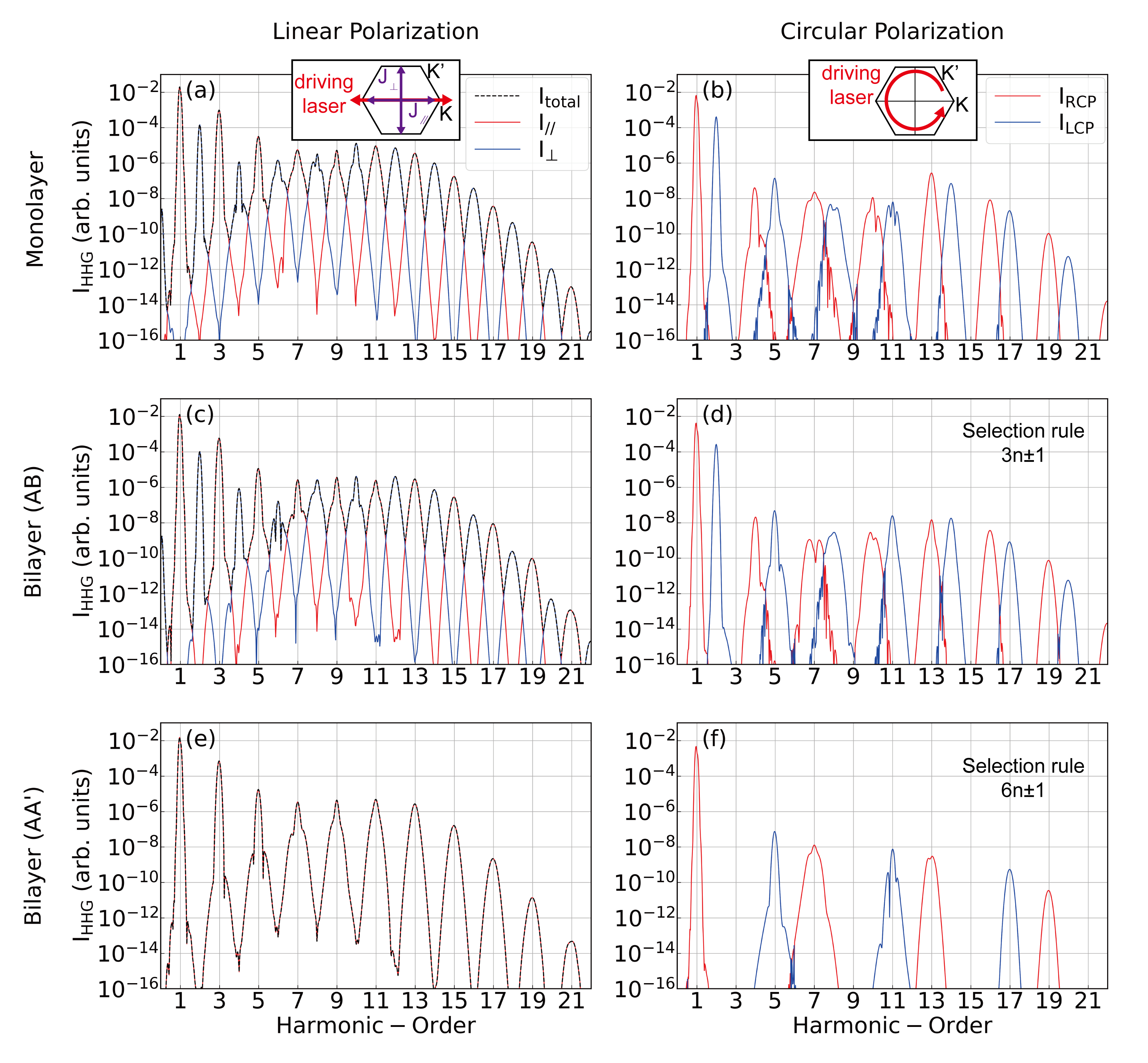}
    \caption{HHG spectra from the monolayer and bilayer hBNs. The HHG spectra generated by (a, c, e)  linearly polarized laser and (b, d, f) circularly polarized laser are shown for (a, b) monolayer, (c, d) AB bilayer and (e, f) AA$'$ bilayer hBN. The insets in (a) and (b) indicate polarization directions of the driving laser, parallel and perpendicular components with respect to the BZ. The driving laser has 1.2$\times \rm 10^{12}$ W/$\rm cm^2$ intensity at 1.9 um wavelength. The carrier-envelope phase is fixed at zero and a gaussian pulse with a FWHM of 10 cycles pulse is considered. The interlayer coupling strength, $t_{11}$, is considered as 0.60 eV (AB) and 0.32 eV (AA$'$) for the case of bilayer hBNs. The laser polarization is along the $\Gamma-K$ direction (x-direction) for the linear case and a right circular polarized driving field was used for circular polarization. The HHG intensities are normalized to the number of layers. 
    }
    \label{fig:HHG_spectra}
\end{figure}

We start by analyzing the effect of layer stacking and symmetry change on the HHG spectra in monolayer and bilayer hBNs. For a 10-cycle  driving field of 1.2$\times \rm 10^{12}$ W/$\rm cm^2$ intensity at 1.9 $\mu$m wavelength, the calculated HHG spectra are displayed for linearly polarized and circularly polarized driving laser fields in Fig.~\ref{fig:HHG_spectra}. The intensities of HHG spectra are normalized to the number of layers in the system. In Fig.~\ref{fig:HHG_spectra}(a, c, e), the HHG spectra show symmetry effects as a function of layer stacking, i.e., monolayer, AB and AA' bilayer. We observe remarkable inversion symmetry effects on the selection rules for all optical harmonic orders (HOs) between Fig.~\ref{fig:HHG_spectra}(a, c) and Fig.~\ref{fig:HHG_spectra}(e).
The inversion symmetry of the AA$'$ bilayer allows us to observe only odd HOs independent of the driving laser field direction (Fig.~\ref{fig:HHG_spectra}(e)). 
For the monolayer and AB bilayer, the $\Gamma-M$ direction is a mirror plane and thus even HOs vanish. When the polarization of the laser is along the $\Gamma-K$ direction, even HOs appear and their directions are perpendicular to the driving laser field (see Fig.~\ref{fig:HHG_spectra}(a). 
In Fig.~\ref{fig:HHG_spectra}(b, d, f), we show the HHG spectra from a right circularly polarized (RCP) driving field.  
Here, the harmonic emissions are analyzed as a sum of the left circular polarization (LCP) and RCP components. Depending on the harmonic order, one of the RCP or LCP components dominates the harmonic emissions. The HHG signal even disappears in a certain HO. The monolayer and AB bilayer hBNs follow the 3n$\pm$1 selection rule. When HO is 3n-1, HHG have opposite circular polarization to the driving laser field (counter-rotating order), and when HO is 3n+1, the HHG have the same circular polarization to the input beam (co-rotating order). There is no signal with the HO 3n. For the case of AA$'$ bilayer hBN, even order harmonics are canceled due to the effect of non-vanishing inversion symmetry. As a result, it follows the 6n$\pm$1 selection rule, which is similar to the graphene case~\cite{chen2019}. 

The basic selection rules come from the symmetry of the system, but there are some minor details that cannot be explained by the symmetry. 
If the interlayer coupling is not considered, the HHG signal from the bilayer system can be simply calculated as a sum of two HHG signals from each band. Then, the symmetry can lead constructive interference in the HHG signal of the AB bilayer as shown in Figures~\ref{fig:HHG_spectra}(c, d). For the case of the AA’ bilayer, the odd order harmonics show constructive interference, while the even order harmonics are canceled by destructive interference (see Figure~\ref{fig:HHG_spectra}(e, f)). 
However, this simple constructive (or destructive) interference cannot perfectly explain the HHG behavior. There are slight differences between the HHG spectra from the monolayer, AB bilayer and AA’ bilayer. The interlayer coupling needs to be considered to explain these differences. 

However, Figure~\ref{fig:HHG_spectra} only shows the HHG signal induced by linearly (along the x-axis in Figure~\ref{fig:material_structure}(d)) and circularly polarized driving fields. To have a better understanding, we analyze the HHG spectra with respect to the polarization angle, ellipticity of driving laser, and interlayer coupling parameter in the following subsections.

\subsection{Polarization angle analysis in HHG spectrum}

\begin{figure}[H]
    \centering
    \includegraphics[width=14cm]{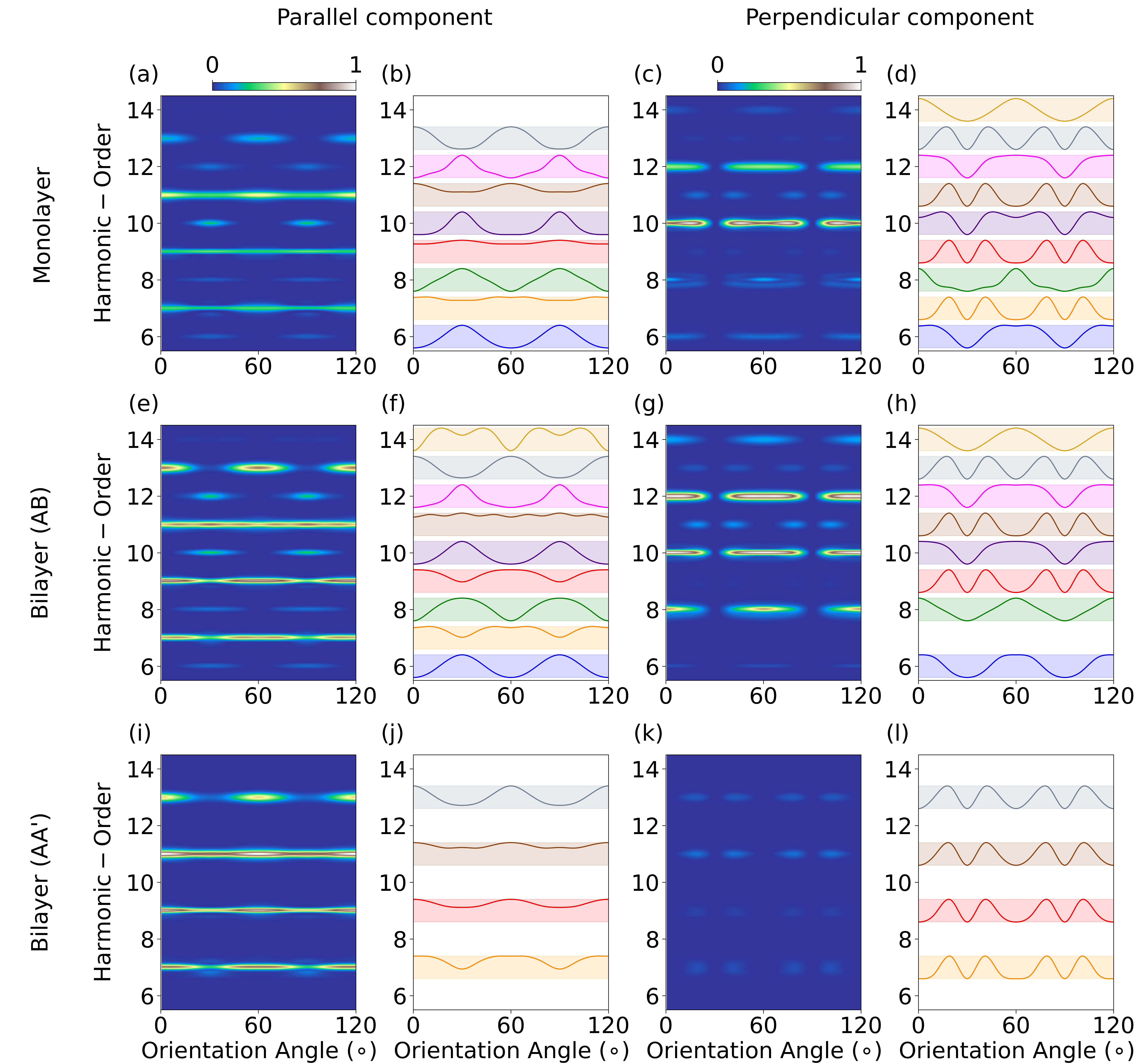}
    \caption{Angular rotation analysis:
    (a) and (c) Color representation of the parallel and perpendicular components, respectively, of HHG spectra depending on the polarization angle obtained from monolayer hBN. The $\Gamma-K$ direction is set to be zero degree. (b) and (d) Normalized high harmonic intensities for parallel and perpendicular components, respectively. (e-h) and (i-l) are the same as (a-d), but from AB and AA$'$ bilayer hBNs.
    }
    \label{fig:angle_scan}
\end{figure}

Figure~\ref{fig:angle_scan} shows the dependence of the HHG spectrum on the laser polarization angle which illustrates: ({\it 1}) the effect of the symmetry and ({\it 2}) the effect of the interlayer coupling by varying laser polarization angles. We analyze the parallel and perpendicular components of HHG to the laser polarization angles of 0$^{\circ}$ to 120$^{\circ}$. Here, we set the laser polarization angle to be 0$^{\circ}$ when it is along the $\Gamma-K$ direction in BZ (see Fig. \ref{fig:material_structure}(d)). The results from three geometrically different types of hBN share many features. Every harmonic order has periodicity of 60$^{\circ}$ which comes from the 6-fold symmetry of the band structure, even though the crystal symmetry is 3-fold. The perpendicular components of the odd HOs spectra are much smaller than those of other HOs. The perpendicular components of the even HOs have maxima in the $\Gamma-K$ direction, while the parallel components have maxima in the $\Gamma-M$ direction for both even and odd HOs. Also, some parallel components of odd harmonics span over the whole angle and there is no clear minima. All of these observations agree with previous works such as the monolayer and AA$'$ bilayer hBN calculation \cite{lebreton2018} or monolayer MoS$_{2}$ (the same symmetry as monolayer hBN) experiment \cite{Liu2017}. 

Although HHG spectra from geometrically different hBNs have some similar tendencies, there are differences due to the change of symmetry and interlayer coupling. 
In the perpendicular components, there is no noticeable difference between the monolayer, AB bilayer and AA$'$ bilayer hBNs. The only difference is that the even HOs in the AA$'$ bilayer are missing due to IS. 
The parallel components show similar structures, but not strictly as in the perpendicular case. For example, the parallel component of HO9 in the monolayer has maximum intensity at 30$^{\circ}$, but it has the minimum intensity in the AB and AA$'$ bilayers. It is interesting that the AB bilayer is closer to the AA$'$ bilayer rather than the monolayer in terms of patterns in parallel components of the HHG, even though the symmetries of the AB and AA$'$ bilayers are different. This result suggests that the perpendicular components heavily depend on crystal symmetry, while the parallel components are more affected by energy dispersion and other terms such as interlayer coupling. 

\subsection{Ellipticity dependence of HHG spectrum}

\begin{figure}[H]
    \centering
    \includegraphics[width=14cm]{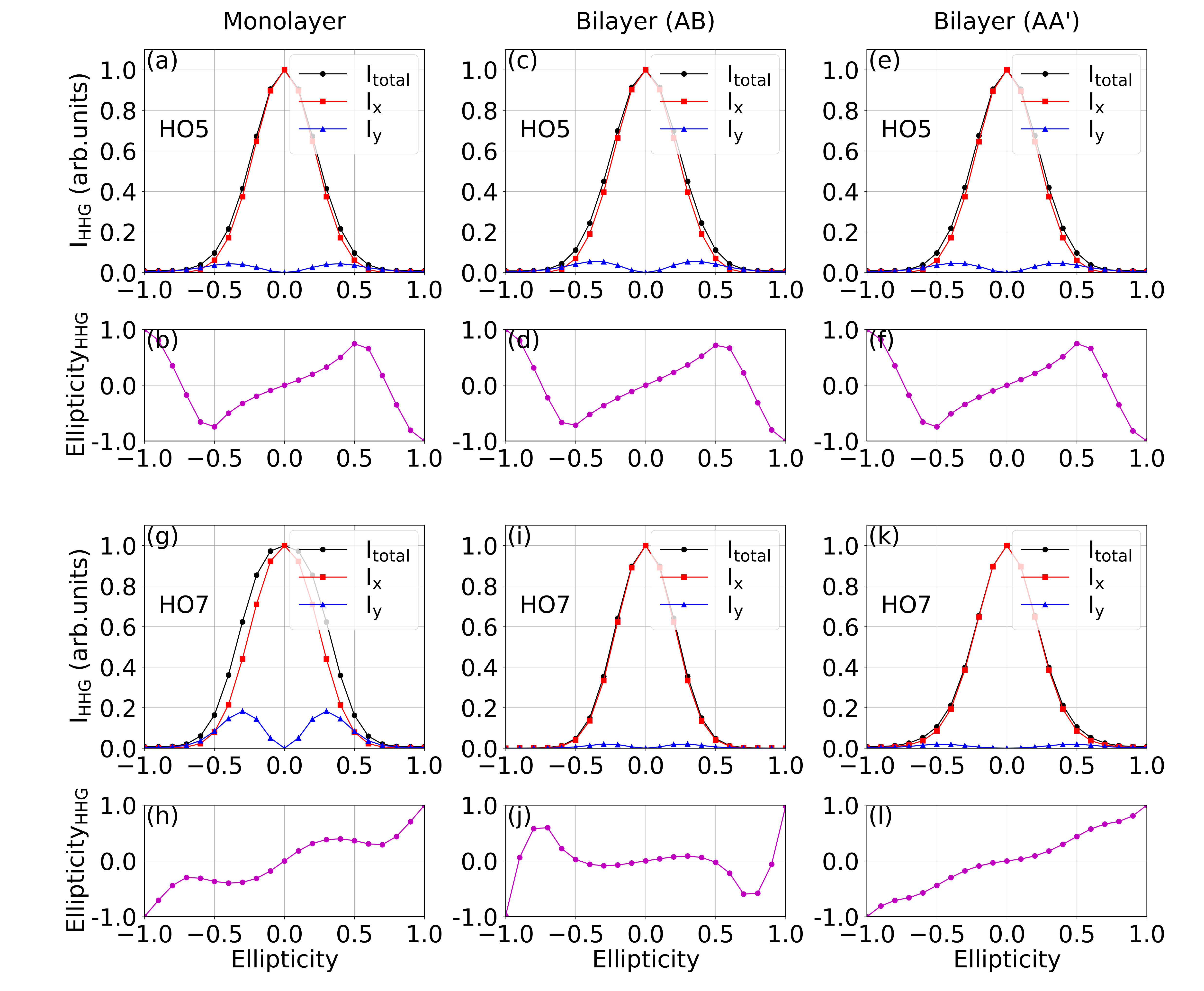}
    \caption{Ellipticity dependence of HHG intensities and ellipticities of the HO5 in (a, b) the monolayer, (c, d) AB bilayer and (e, f) AA$'$ bilayer hBNs. The top panels (a,c,e) show HHG intensities of HO5 as a function of ellipticity. The bottom panels (b,d,f) show ellipticity of HO5 as a function of driving laser ellipticity. Normalized HHG intensity, x components and y components are shown in black, red and blue lines, respectively. The major axis of the elliptically polarized laser field is along the $\Gamma-K$(x) direction. The right (left) circular polarization is set to be 1 (-1) ellipticity. The same analysis for the HO7 is shown in (g-l). 
    }
    \label{fig:ellipticity_scan}
\end{figure}

The ellipticity dependence of HHG emission is shown in Figure~\ref{fig:ellipticity_scan} for monolayer and bilayer hBNs. We use the laser defined by Eq.~\ref{eq:vecotr_potential}, and change the ellipticity of the driving laser from -1 (left circular polarization) to 0 (linear polarization) to 1 (right circular polarization). The major axis of the elliptical polarization or linear polarization direction is set to be in the $\Gamma-K$(x) direction. We analyze the intensity of the HHG spectrum in HO5 and HO7, and also show the ellipticity of the output HHG signal as a function of the ellipticity of the driving laser field. Since HHG emission from the AA$'$ bilayer hBN is limited by the 6n$\pm$1 selection rule for a circularly polarized driving field, we will focus on the HO5 and HO7 for our study. We already observed that HO5 and HO7 have a constructive interference for both bilayers in Figure~\ref{fig:HHG_spectra}(b, d, f). Figure~\ref{fig:ellipticity_scan} shows that they also have a constructive interference in HO5 and HO7 with the elliptically polarized laser. For the case of HO5, the ellipticity dependencies of the HHG spectra in monolayer and bilayer cases are almost similar both in intensity (see Figure~\ref{fig:ellipticity_scan}(a, c, e)) and ellipticity analysis (see Figure~\ref{fig:ellipticity_scan}(b, d, f)). They have a similar and typical pattern - maximized at linear polarization and minimized at circular polarization. It is impressive that even the tendencies of ellipticity also match each other very well (see Figures~\ref{fig:ellipticity_scan}(b, d, f)). 
For the case of HO7, the total intensity is similar (see Figures~\ref{fig:ellipticity_scan}(g, i, k)), but there are some differences in the ellipticity of the output beam (see Figures~\ref{fig:ellipticity_scan}(h, j, l)). There is a local maximum of output ellipticity in the monolayer case. The AB bilayer shows the enhancement of the local peak, while the AA$'$ bilayer cancels the peak and flattens the response. It is interesting that although the strength of the local minimum is different, the location of the peak is the same in each case. HO11 and HO13 also behave similarly. It is clear that the output beam ellipticity response is very sensitive to changes in material structure. However, there are some preserved properties that might come from the lattice structure.

\subsection{Effect of interlayer coupling on HHG spectrum}

\begin{figure}
    \centering
    \includegraphics[width=14cm]{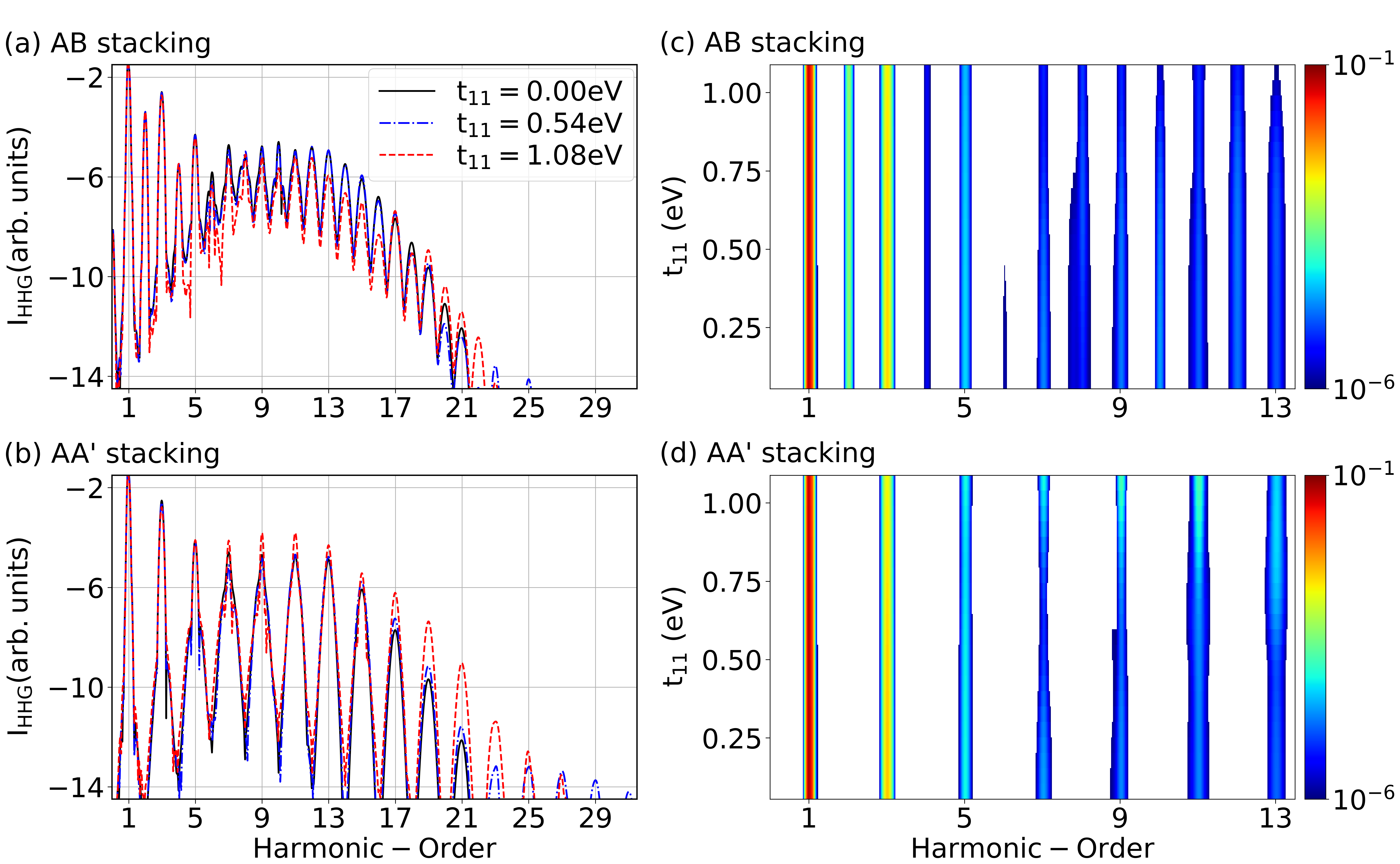}
    \caption{High-order harmonic spectra of (a) AB and (b) AA$'$ bilayer hBN for three different interlayer coupling strengths, $t_{11}$=0 eV, $t_{11}$=0.54 eV and $t_{11}$=1.08 eV. The color representation of HHG intensities as a function of inter layer coupling is displayed in (c) and (d) for AB and AA bilayer hBNs, respectively.}
    \label{fig:t11_scan}
\end{figure}

Figure~\ref{fig:t11_scan} presents the effect of interlayer coupling ($t_{11}$) on the HHG spectrum. We plot the HHG spectra as a function of three $t_{11}$ values ($t_{11}$ = 0 eV, 0.54 eV, 1.08 eV) for the AB (Figure~\ref{fig:t11_scan}a) and AA$'$(Figure~\ref{fig:t11_scan}b) bilayers. We observe that high HOs (HOs after the first plateau) get stronger as $t_{11}$ increases. Although there was an observation of amplification of HHG intensities in previous experiments for graphene/hBN hetero structure \cite{chen2020}, their explanation cannot be applied here. In Ref.~\cite{chen2020}, the enhancement is only on the perpendicular components of the HHG signal in all HOs. In this study, only the cutoff region shows clear enhancement, while the plateau region remains complex. We divide the HHG spectrum into low-order and high-order region. When the strong interlayer coupling is applied, the additional transition path become stronger and the band splitting becomes larger. Additionally, the energy gaps are also larger as $t_{11}$ increases. A larger maximum band gap and multiple paths lead to the enhancement of HHG signal in the high HOs.     

To see the change in low HOs in detail, we scan through $t_{11}$ values in Figure~\ref{fig:t11_scan}(c, d). The effect of $t_{11}$ is straight-forward in the AB bilayer. Although the variation is small, the HHG intensities have a monotonically decreasing tendency as the $t_{11}$ value increases. In the case of the AA$'$ bilayer, the change is more complex. When the $t_{11}$ value increases, the HHG intensities decrease in the beginning, but at some point, they start to increase. In the end, stronger HHG spectrum is produced in the AA$'$ bilayer if $t_{11}$ is sufficiently increased. This can be understood through the analysis of dipole transition matrix elements (DTMEs). The absolute values of the DTMEs between the highest valence band (VB2) and two conduction bands (CBs) are plotted for the AB (Figure~\ref{fig:dipole_AB}) and AA$'$ (Figure~\ref{fig:dipole_AAp}) bilayers. 

\begin{figure}
    \centering
    \includegraphics[width=14cm]{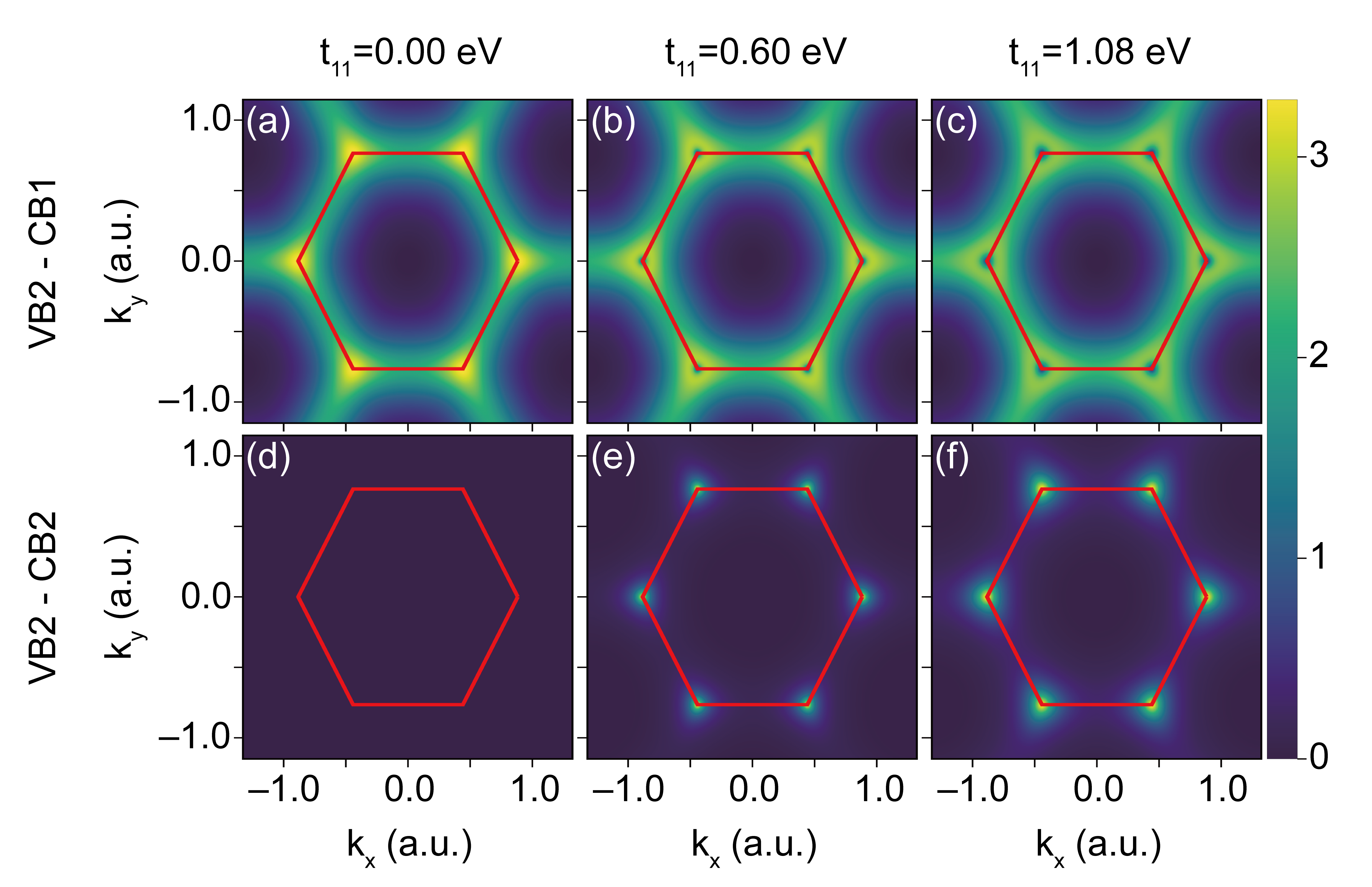}
    \caption{Absolute value of dipole transition matrix elements in AB bilayer. (a-c) show the dipole transition matrix elements between highest valence band VB2 and lowest conduction band CB1 ($\bm{d}_{v_{2},c_{1}}$) for different interlayer coupling strength $t_{11}$= 0 eV, $t_{11}$=0.60 eV and $t_{11}$=1.08 eV, respectively. The red-colored hexagon indicates the BZ. The dipole transition matrix elements between highest valence band VB2 and highest conduction band CB2 ($\bm{d}_{v_{2},c_{2}}$) are displayed in (d-f).}
    \label{fig:dipole_AB}
\end{figure}

In the AB bilayer, we observe that the DTMEs between CB2 and VB2 ($\bm{d}_{v_{2},c_{2}}$) are nearly zero without interlayer coupling ($t_{11}\approx$0) as shown in Figure~\ref{fig:dipole_AB}(d)). The bilayer can be treated as two independent layers without interlayer coupling. Therefore, the existing DTMEs in Figure~\ref{fig:dipole_AB}(a) indicate that VB2 band and CB1 belong to one layer, while VB1 and CB2 belong to the other layer.   
In addition, the DTMEs between VB2 and CB1 seem to be transferred to the DTMEs between VB2 and CB2 during the interlayer coupling gets stronger. This explains monotonic change shown in Figure~\ref{fig:t11_scan}.

\begin{figure}
    \centering
    \includegraphics[width=14cm]{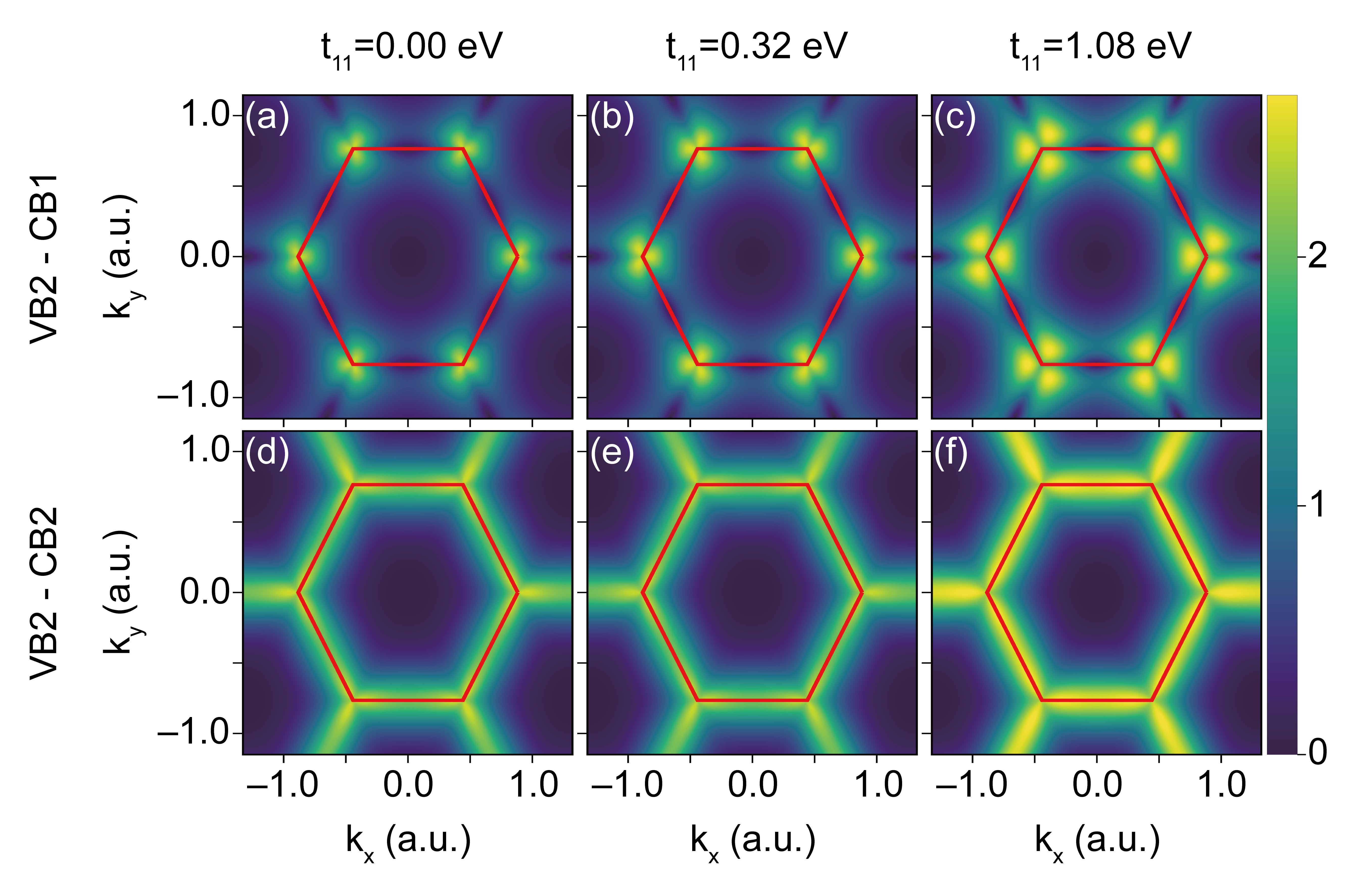}
    \caption{Absolute value of dipole transition matrix elements in AA$'$ bilayer. The dipole transition matrix elements are displayed as in Fig.~\ref{fig:dipole_AB}.}
    \label{fig:dipole_AAp}
\end{figure}

In the case of the AA$'$ bilayer, different behavior is observed. As shown in Figure~\ref{fig:dipole_AAp}(a, d) corresponding to an extremely small $t_{11}$ value ($t_{11}$ $\approx$ 3.7$\times$10$^{-9}$ eV), the shape and strength of the DTMEs are already similar to those of equilibrium ($t_{11}$ = 0.32 eV) Figure~\ref{fig:dipole_AAp}(b,e). This implies that bands from different layers are already mixed even if the $t_{11}$ is small. If the bands from two layers are already mixed, the increase in $t_{11}$ does not simply mean the increase in interaction strength between the bands from different layers, but the strength and shape of the interaction has a complex modification. This can also be predicted by energy dispersion degeneracy at $\bm{K}$ point (Figure~\ref{fig:material_structure}(f)). This degeneracy cannot be avoided because it comes from the structure of the Hamiltonian, and directly indicates the mixing of the bands from two layers.



\section{Conclusion}\label{sec:conclustion}

In summary, we study the effect of interlayer coupling and crystal symmetry with respect to laser polarization angle, driving laser ellipticity, and interlayer coupling $t_{11}$. 
\begin{itemize}
\item The HHG spectrum from the AB or AA$'$ bilayer can be approximately understood as the constructive/destructive interference of two independent monolayer HHG spectra.
\item The effect of interlayer coupling is well manifested in parallel odd components.
\item Which aspects are affected by the interlayer coupling are specified.
\item More importantly, the effect of interlayer coupling on HHG intensity is analyzed and explained via dipole analysis. 
\end{itemize}
 

We believe this work will help in the understanding of the interlayer mechanics of bilayer materials, and offer a guide to possible experiments. The role of the interlayer coupling and changes in symmetry could be extended to investigate or design materials from a two-dimensional monolayer to a multiple layer system such as van der waals heterostructure. 

\vspace{6pt} 



\authorcontributions{
Conceptualization, Dasol Kim; Methodology, Dasol Kim and Alexis Chac\'{o}n; software, Dasol kim and Alexis Chac\'{o}n; Formal analysis, Dasol Kim and Yeon Lee; Investigation, Dasol Kim; Resources, Dong Eon Kim; Data curation, Yeon Lee; writing-original draft, Dasol Kim and Yeon Lee; writing-review \& editing, Dong Eon Kim; visualization, Yeon Lee; Supervision, Alexis Chac\'{o}n; validation, Dasol Kim; project administration, Dong Eon Kim; funding acquisition, Dong Eon Kim. Dasol Kim and Yeon Lee contributed equally as first authors. All authors have read and agreed to the published version of the manuscript.
}

\funding{The work has been supported in part by Grant No 2016K1A4A4A01922028 (the Max Planck POSTECH/KOREA Research Initiative Program), Grant No 2020R1A2C2103181 through the National Research Foundation of Korea (NRF) funded by the Ministry of Science and ICT, and Korea Institute for Advancement of Technology (KIAT) grant (No. P0008763, The Competency Development Program for Industry Specialists) funded by MOTIE.}

\institutionalreview{Not applicable}

\informedconsent{Not applicable}


\dataavailability{Not applicable}

\acknowledgments{We thank the Max Planck Institute for the Structural Dynamics of matter (MPI-SD) in Hamburg, Germany (Prof. Angel Rubio) for providing generous computational resources for the intense calculation.}

\conflictsofinterest{The authors declare no conflict of interest. The funders had no role in the design of the study; in the collection, analyses, or interpretation of data; in the writing of the manuscript, or in the decision to publish the~results.} 

\sampleavailability{Not applicable}

\end{paracol}
\reftitle{References}


\externalbibliography{yes}
\bibliography{hBN}

%


\end{document}